\documentclass[twocolumn,showpacs,preprintnumbers,amsmath,amssymb]{revtex4}
\usepackage{graphicx}
\usepackage{dcolumn}
\usepackage{bm}
\begin{document}
\title{Remote-doping scattering and the local field corrections in
the 2D electron system in a modulation-doped Si/SiGe quantum well}
\author{V.~T. Dolgopolov, E.~V. Deviatov, and A.~A. Shashkin}
\affiliation{Institute of Solid State Physics, Chernogolovka, Moscow
District 142432, Russia}
\author{U. Wieser and U. Kunze}
\affiliation{Ruhr-Universit\"at Bochum, D-44780 Bochum, Germany}
\author{G. Abstreiter and K. Brunner}
\affiliation{Walter Schottky Institute, Technische Universit\"at
M\"unchen, W-8046, Garching, Germany}
\begin{abstract}
The small, about 30\% magnetoresistance at the onset of full spin
polarization in the 2D electron system in a modulation-doped Si/SiGe
quantum well gives evidence that it is the remote doping that
determines the transport scattering time. Measurements of the
mobility in this strongly-interacting electron system with
remote-doping scattering allow us to arrive at a conclusion that the
Hubbard form underestimates the local field corrections by about a
factor of 2.
\end{abstract}
\pacs{73.43.-f, 72.20.-i, 72.20.Ee}
\maketitle

Much interest has been attracted recently by the behavior of
two-dimensional (2D) electron systems in a parallel magnetic field.
The resistance of a 2D electron system in Si MOSFETs was found to
rise with parallel field saturating to a constant value above a
critical magnetic field $B_c$ \cite{sat} which corresponds to the
onset of full spin polarization of the electron system \cite{okvit}.
The spin origin of the effect is consistent with its insensitivity to
the direction of the in-plane field \cite{sat}. At low electron
densities, the experimental resistance ratio is equal to
$R(B_c)/R(0)\approx 4$ as long as the system remains metallic
\cite{remark}, which is in agreement with the calculation \cite{dol}.
The effect was used to study the spin susceptibility of the
least-disordered electron system \cite{sh} as well as the local
moments in the bandtail in more-disordered Si MOSFETs
\cite{pud,broto}. In contrast, the 2D carrier system in GaAs/AlGaAs
heterostructures is relatively thick so that the orbital effects
become important and give rise to an enhancement of the effective
mass in parallel magnetic fields \cite{zhu,tutuc}. As a result, in
GaAs there are two noteworthy distinctions \cite{simmons}: (i) above
$B_c$, the resistance keeps on increasing less steeply with no sign
of saturation; and (ii) the magnetoresistance is strongly anisotropic
depending upon the relative orientation of the in-plane magnetic
field and the current.

The 2D carrier systems in Si MOSFETs and high-mobility GaAs/AlGaAs
heterostructures are similar in that at low carrier densities, the
transport scattering time is determined by charged impurities near
the 2D system. In the former system charged impurities are located at
the Si/SiO$_2$ interface resulting in dominant large-angle scattering
\cite{ando}, whereas in the latter they are homogeneous background
doping for both 2D electrons \cite{apl} and 2D holes \cite{prosk}.
The case of a 2D carrier system with a finite spacer that is
remarkable by remote-doping scattering, i.e., dominant small-angle
scattering, is opposite. Although it may seem simple to realize such
a 2D system, that kind of scattering has not been unequivocally
established in any 2D system studied so far. Recently, it has been
predicted that for remote doping, the above resistance ratio should
be equal to $R(B_c)/R(0)\approx 1.2$ \cite{gold}.

In this paper, we report measurements of the resistance of the 2D
electron system in a modulation-doped Si/SiGe quantum well in
parallel magnetic fields. Being very similar to (100)-Si MOSFETs,
this electron system is different by the presence of a spacer and the
1.5 times larger dielectric constant so that the same strength of
electron-electron interactions in Si/SiGe can be expected at 2.2
times lower electron densities. The fact that the observed
magnetoresistance is small shows that the transport scattering time
in our samples is determined by remote doping. In this regime, the
magnetoresistance is sensitive to a small amount of the charged
residual impurities at the Si/SiGe interface ($< 1$\% of the electron
density), which varies in different runs depending on external
perturbations such as cooling and illumination of the sample. Using
the dependence of the mobility on electron density in this
strongly-interacting system with remote-doping scattering, we extract
the form of the local field corrections (LFC). From comparison of the
experimental data and the model calculations it follows that account
should be taken of both kinematic and dynamic correlations.

Samples were grown by molecular beam epitaxy on (001)-Si substrates.
A 15~nm thick Si channel was deposited on a strain-relaxed buffer
layer which consists of a 600~nm Si$_{0.7}$Ge$_{0.3}$ buffer on the
top of a 2.5~$\mu$m thick Si$_{1-x}$Ge$_x$ layer with compositional
grading. The channel was capped by 14~nm Si$_{0.7}$Ge$_{0.3}$ spacer,
12~nm P-doped layer, and 27.5~nm cap layer, covered by 10~nm of Si.
The samples were arranged in a standard Hall bar geometry. The
channel of the transistor had a uniform width of 20~$\mu$m between
the ohmic source and drain contacts. The four voltage probes and the
source/drain electrode were formed by local implantation of
phosphorous ions. Subsequent activation and recrystallization was
performed at 560$^\circ$~C for 30~min in an N$_2$ ambient. Afterwards
the contact pads were metallized with 25~nm Ti and 100~nm Au. In
order to reduce the contact resistance, the sample was annealed at
400$^\circ$~C for 60~s in an N$_2$ atmosphere. Finally, the Schottky
gate electrode comprising 30/100~nm Ni/Au was defined by a lift off
process.

\begin{figure}\vspace{-0.15in}
\scalebox{0.36}{\includegraphics{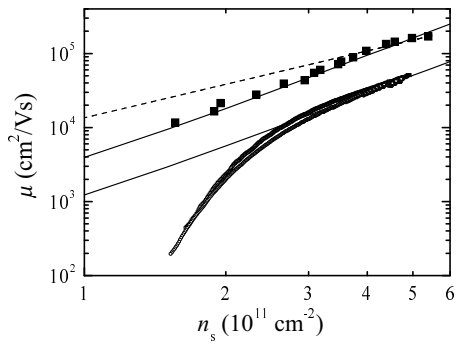}}\vspace{-2.1in}
\caption{\label{wil12} Zero-field mobility as a function of electron
density at a temperature of 30~mK for two samples (open symbols)
along with the data of Ref.~\cite{wil} (filled squares). Also shown
is the result of the calculation with (solid lines) and without LFC
(dashed line) neglecting the numerical factor.}
\end{figure}

The sample was placed in the mixing chamber of a dilution
refrigerator with a base temperature of 30~mK. To create the mobile
carriers in the 2D system, the sample was illuminated with a
light-emitting diode until the resistance saturated. After the diode
was switched off, the sample state did not change in the run. The
resistance was measured using a standard four-terminal lock-in
technique at a frequency of 15~Hz in magnetic fields up to 14~T.
Excitation current through the device was kept low enough ($< 10$~nA)
to ensure that measurements were taken in the linear regime of
response. To change the sample position in the mixing chamber we
warmed the sample up, rotated it at room temperature, and cooled down
again. The alignment uncertainty of the sample plane with the
magnetic field was kept within $0.3^\circ$. The electron density as a
function of gate voltage was determined from Shubnikov-de Haas
oscillations in perpendicular magnetic fields. We have verified that
the gate voltage dependence of the resistance in zero magnetic field
is well reproducible in different runs with the accuracy of
insignificant threshold shifts. In parallel magnetic fields, this
dependence was used for determining the threshold voltage.

In Fig.~\ref{wil12}, we show the dependence of the zero-field
mobility, $\mu$, on electron density, $n_s$, for two of our samples
along with the data of Ref.~\cite{wil} obtained on a Si/SiGe quantum
well with higher mobility. We have verified that in the studied range
of electron densities, the mobility is temperature-independent below
1~K. Despite the difference between the mobilities is quite
appreciable, the slopes of all dependences above $3\times
10^{11}$~cm$^{-2}$ are coincident and correspond to a power law:
$\mu\propto n_s^{2.4}$, as was observed previously \cite{wil,rem1}.
Below $3\times 10^{11}$~cm$^{-2}$, the stronger decrease of the
mobility with lowering electron density in our samples is likely to
be a precursor of Anderson localization as caused by multiple
scattering \cite{agold}.

\begin{figure}\vspace{-0.15in}
\scalebox{0.36}{\includegraphics{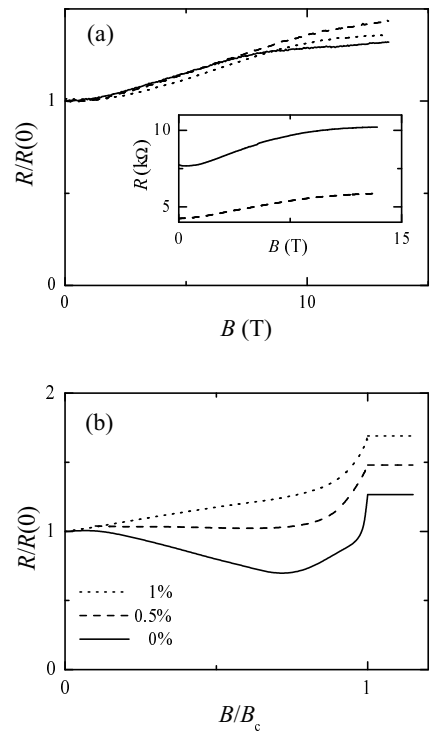}}\vspace{-0.05in}
\caption{\label{Rsat} (a) The normalized magnetoresistance measured
in different runs at $n_s=2.98\times 10^{11}$~cm$^{-2}$ and at a
temperature of 30~mK for $B\parallel I$ (solid and dotted lines) and
$B\perp I$ (dashed line). The inset shows the magnetoresistance on
another sample in one run at $n_s=2.34\times 10^{11}$ and $2.54\times
10^{11}$~cm$^{-2}$ for $B\perp I$. (b) The calculated
magnetoresistance at $n_s=2\times 10^{11}$~cm$^{-2}$ for different
parts of the charged impurities at the interface.}
\end{figure}

In Fig.~\ref{Rsat}(a), we show the magnetoresistance of the sample in
parallel fields. As the magnetic field is increased, the resistance
increases and tends to saturate at the onset of full spin
polarization in this electron system. The magnetoresistance is
practically independent of the relative orientation of the field and
the current, $I$. In other words, no anisotropy of the resistance
with respect to the in-plane field is observed in our samples, which
is similar to the case of Si MOSFETs. However, as compared to Si
MOSFETs, the ratio $R(B_c)/R(0)$ in the studied range of electron
densities is much smaller, about 1.3 \cite{rem}. This indicates
directly that the remote-doping scattering prevails \cite{gold}. The
form of the $R(B)$ curve varies slightly in different runs. Even in
the same run, it changes slightly with changing electron density
(inset to Fig.~\ref{Rsat}).

\begin{figure}\vspace{-0.15in}
\scalebox{0.36}{\includegraphics{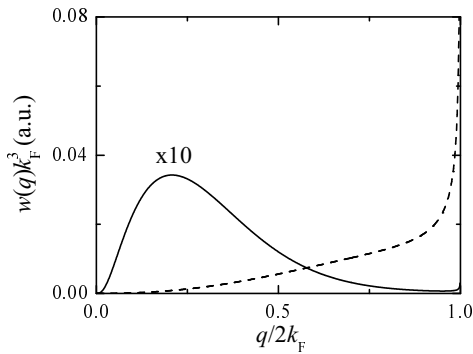}}\vspace{-2.1in}
\caption{\label{wil13} Dependence of the integrand in Eq.~(\ref{eq1})
on transferred wavevector for $n_s=10^{11}$~cm$^{-2}$ (dashed line)
and $n_s=10^{12}$~cm$^{-2}$ (solid line) assuming $G(q)=0$.}
\end{figure}

We now discuss the regime of remote-doping scattering. The screening
properties of a 2D electron system are determined by two parameters:
the screening wavevector, $q_s=2g_vme^2/\varepsilon\hbar^2$, and the
Fermi wavevector, $k_F=(2\pi n_s/g_v)^{1/2}$ (where $g_v$ is the
valley degeneracy, $m$ is the band mass, and $\varepsilon$ is the
dielectric constant). For our case the former equal to
$q_s=1.25\times 10^7$~cm$^{-1}$ exceeds the latter which falls within
the range $1.4\times 10^6$~cm$^{-1} <2k_F< 2.5\times 10^6$~cm$^{-1}$
corresponding to the Wigner-Seitz radius $4.5 >r_s> 2.5$. The third
parameter that controls the electron scattering in Si/SiGe is half
the inverse spacer width, $1/2d$, which is equal to $1.4\times
10^6$~cm$^{-1}$ in our samples. As in the high-$n_s$ limit the
transferred wavevector $1/2d$ is small compared to $2k_F$, the
electron backscattering is small. In the opposite limit the
transferred wavevector $1/2d$ is approximately equal to $2k_F$ and,
therefore, the electron backscattering should occur. Nevertheless, we
argue that this is still dominant small-angle scattering. Indeed, in
the lowest order of multiple scattering theory \cite{gotze} the
inverse transport scattering time for the two-valley case can be
written

\begin{equation}
\frac{1}{\tau}=\frac{m}{\pi\hbar^3}\int_0^{2k_F}\frac{q^2dq}
{k_F^2\sqrt{4k_F^2-q^2}}\frac{V_q^2}{[1+\frac{q_s}{q}(1-G(q))]^2},
\label{eq1}
\end{equation}
where $G(q)$ is the LFC and $V_q^2$ in the case of remote doping is
given by

\begin{equation}
V_q^2=\frac{(2\pi e^2)^2}{\varepsilon^2q^2}\int_d^{d_1}N_i(z)
\exp(-2qz)dz.\label{eq2}
\end{equation}
Here $d_1-d$ is the width of the doped layer with doping density
$N_i$. The dependence of the integrand, $w(q)$, in Eq.~(\ref{eq1}) on
transferred wavevector at high and low electron densities is
displayed in Fig.~\ref{wil13} ignoring the LFC. In both limits the
main contribution to the scattering probability originates from
wavevectors well below $2k_F$, in contrast to Si MOSFETs where the
close vicinity of $2k_F$ contributes only.

\begin{figure}\vspace{-0.15in}
\scalebox{0.36}{\includegraphics{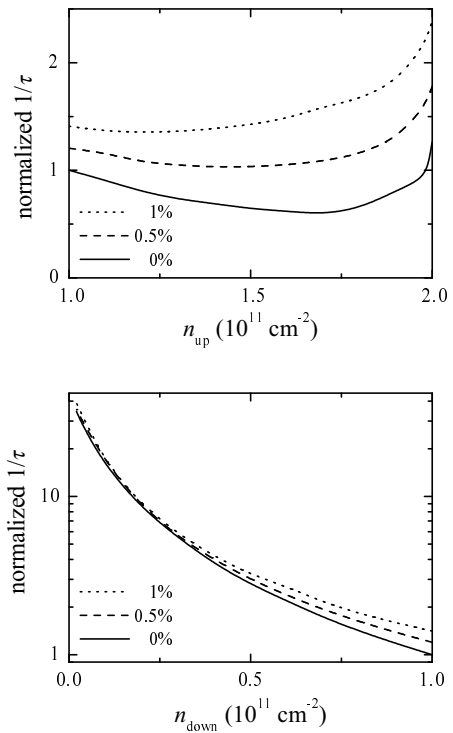}}\vspace{-0.05in}
\caption{\label{tau} The calculated scattering rate for spin-up and
spin-down electrons vs corresponding electron density at $n_s=2\times
10^{11}$~cm$^{-2}$ for different parts of the charged impurities at
the interface. All dependences are normalized by the scattering rate
for remote doping in zero magnetic field.}
\end{figure}

It is remarkable that in the regime of remote-doping scattering, the
magnetoresistance at the onset of complete spin polarization is small
\cite{gold}. A qualitative account of this effect is given below.
Assuming that the spin flip processes are absent, the calculated
scattering rate $1/\tau$ for spin-up and spin-down electrons as a
function of corresponding electron density is shown in
Fig.~\ref{tau}. With increasing $n_{\text{up}}$ (or
$n_{\text{down}}$) the scattering angle decreases and, therefore, the
scattering rate decreases except at the onset of complete spin
polarization at which the increase of the scattering rate is related
to the change of screening, similar to the effect discussed in
Ref.~\cite{dol}. The resulting magnetoresistance is negative and only
becomes positive near $B_c$, the ratio $R(B_c)/R(0)$ being equal to
1.25, see Fig.~\ref{Rsat}(b). It is important that if there is a
small amount of charged impurities at the Si/SiGe interface, the
negative magnetoresistance is suppressed accompanied by somewhat
larger $R(B_c)/R(0)$. The calculated magnetoresistance with 0.5\% of
the charged impurities (about $2\times 10^9$~cm$^{-2}$) located at
the interface describes the experiment reasonably well, see
Fig.~\ref{Rsat}. So, density variations of the charged residual
impurities at the Si/SiGe interface and, hence, of the local moments
in the bandtail \cite{pud} naturally explain the observed changes in
$R(B)$ including those in different runs.

If the local field corrections are still disregarded, the solution to
Eq.~(\ref{eq1}) is $\mu\propto n_s^{3/2}$ \cite{agold} as shown by
the dashed line in Fig.~\ref{wil12}. This dependence is noticeably
weaker than the experimental one. The origin for the discrepancy is
strong correlations in this electron system. At low electron
densities, worse screening is reflected by LFC so that the integrand
in Eq.~(\ref{eq1}) should be larger at wavevectors about $2k_F$. The
Hubbard form of the LFC

\begin{equation}
G(q)=\frac{1}{2g_v}\frac{q}{\sqrt{q^2+k_F^2}}\label{eq3}
\end{equation}
yields a power law $\mu\propto n_s^{1.8}$, the exponent being still
smaller compared to the experimental finding. This is not very
surprising because the Hubbard form includes only kinematic
correlations that are caused by Pauli principle. Another contribution
is given by dynamic correlations that are related to direct Coulomb
interelectron interactions; these contribute to the LFC at large
wavevectors \cite{davoudi}. To simulate the effect of dynamic
correlations, we demand $g_v=1$ in Eq.~(\ref{eq3}). The so-obtained
dependence $\mu(n_s)$ is indicated by the solid lines in
Fig.~\ref{wil12} disregarding the numerical factor. Agreement between
the calculation and the experiment shows that the LFC are
approximately twice as large as the Hubbard form.

Finally, we discuss the above assumption of the absence of spin flip
processes. Available experimental data for Si MOSFETs allows one to
conclude that such processes are in fact present but are not
dominant. Indeed, the Hall resistance in a magnetic field with weak
perpendicular and strong parallel components was found to be
significantly lower than that expected for decoupled spin subbands
\cite{vitk}. At the same time, ratios $R(B_c)/R(0)\approx 4$ were
observed at low electron densities in the metallic regime \cite{sat};
these are close to the ratio $R(B_c)/R(0)=4$ expected for the case
when no spin flip occurs, whereas twice as low a ratio is expected if
the spin flip time is the shortest \cite{dol}. This gives evidence
that the spin flip processes are not of importance and, thus, the
above considerations are justified.

In summary, we have established that in the 2D electron system in a
modulation-doped Si/SiGe quantum well, the regime of remote-doping
scattering occurs in which the parallel-field magnetoresistance is
small and sensitive to uncontrollable density variations of the
charged residual impurities at the Si/SiGe interface. Based on
analysis of the mobility as a function of electron density in this
regime, we conclude that the local field corrections are
approximately double the Hubbard form and, therefore, both kinematic
and dynamic correlations make significant contribution.

We gratefully acknowledge discussions with A. Gold, V.~S. Khrapai,
and S.~V. Kravchenko. We would also like to thank D. Scheible, C.
Meyer, and S. Manus for technical help. This work was supported by
A.~von Humboldt Foundation via Forschungspreis, the Russian
Foundation for Basic Research, and the Russian Ministry of Sciences.





\begin{thebibliography}{apssamp}
\bibitem{remark} At electron densities below approximately $1.5n_c$
(where $n_c$ is the density for the zero-field metal-insulator
transition), the application of a strong parallel magnetic field
makes the system insulating, in which case the ratio $R(B_c)/R(0)$
may become arbitrarily high \cite{sat}.
\bibitem{rem1} An attempt to account for the strong dependence
$\mu(n_s)$ was made in Ref.~\cite{wil} based on the idea of worse
screening of potential fluctuations with lowering $n_s$ due to a
decrease of the average thermodynamic density of states near the
metal-insulator transition. However, as the range of potential
fluctuations found in that paper exceeds the electron mean free path,
the scattering calculations using the average thermodynamic density
of states in such an inhomogeneous electron system are obviously
incorrect.
\end{thebibliography}
\end{document}